\documentclass[11pt,onecolumn,a4paper]{article}
\usepackage{amsmath}
\usepackage{amssymb}
\usepackage{graphicx}
\usepackage{bm}
\usepackage{tikz}
\usetikzlibrary{automata, arrows, positioning, calc, bending, decorations.text, decorations.pathreplacing, angles, quotes, fit, patterns}
\usepackage[utf8]{inputenc}
\usepackage{chemformula}
\usepackage{epsfig}
\usepackage{url}
\usepackage{enumitem}
\usepackage{subfig}
\usepackage{epstopdf}
\usepackage{svg}
\usepackage{forest}
\usepackage{algorithm}
\usepackage[noend]{algpseudocode}
\usepackage{booktabs}
\usepackage{pgfplots}
\usepackage{siunitx}
\usepackage{soul}
\usepackage[hidelinks]{hyperref}
\usetikzlibrary{decorations.pathreplacing}
\tikzset{
  font=\normalsize,
  red arrow/.style={
    midway,red,sloped,fill, minimum height=1.5cm, single arrow, single arrow head extend=.6cm, single arrow head indent=.25cm,xscale=0.3,yscale=0.15,
    allow upside down
  },
  black arrow/.style 2 args={-stealth, shorten >=#1, shorten <=#2},
  black arrow/.default={1mm}{1mm},
  tree box/.style={draw, rounded corners, inner sep=.3em},
  node box/.style={white, draw=black, text=black, rectangle, rounded corners},
}

\usepackage{caption}
\usepackage{xcolor}
\usepackage{mathtools}

\title{\Large{Classification-based detection and quantification of cross-domain data bias in materials discovery}}
\author{Giovanni Trezza$^{1}$, Eliodoro Chiavazzo$^{1}$\thanks{Corresponding author: eliodoro.chiavazzo@polito.it}\\ \small{\emph{$^{1}$Department of Energy, Politecnico di Torino, C.so Duca degli Abruzzi 24, Torino 10129, Italy}}}
\date{}

\usepackage[left=2cm, right=2cm, top=2cm]{geometry}

\begin{document}

\maketitle

\begin{abstract}
It stands to reason that the amount and the quality of data is of key importance for setting up accurate AI-driven models.
Among others, a fundamental aspect to consider is the bias introduced during sample selection in database generation.
This is particularly relevant when a model is trained on a specialized dataset to predict a property of interest, and then applied to forecast the same property over samples having a completely different genesis. 
Indeed, the resulting biased model will likely produce unreliable predictions for many of those out-of-the-box samples.
Neglecting such an aspect may hinder the AI-based discovery process, even when high quality, sufficiently large and highly reputable data sources are available.
In this regard, with superconducting and thermoelectric materials as two prototypical case studies in the field of energy material discovery, we present and validate a new method (based on a classification strategy) capable of detecting, quantifying and circumventing the presence of cross-domain data bias.

\begin{description}
\item[Keywords] \emph{Machine Learning; Materials Discovery; Data bias; Superconductors; Thermoelectric materials}
\end{description}
\end{abstract}

\section{Introduction\label{sec:level1}}
In the realm of scientific exploration and technological advancement, the use of Artificial Intelligence (AI) has catalyzed breakthroughs across various scientific and technological domains \cite{wang2023scientific}, comprising (and not limiting to) predicting protein structures \cite{jumper2021highly}, solving olympiad geometry problems \cite{trinh2024solving}, learning energy surfaces of many body systems incorporating group theory \cite{han23w}, solving high dimensional partial differential equations \cite{han2018solving}, automating data collection, visualization and processing \cite{akiyama2019first}, aiding in theory formulations \cite{wagner2021constructions}, suggesting experiments to be performed \cite{coley2019robotic}.
One such domain that has witnessed significant transformation is Materials Science \cite{vasudevan2019materials, guo2021artificial, decost2020scientific, lopez2023artificial}, where AI-driven approaches is believed to have the potential to revolutionize the search for novel materials with desired properties \cite{zeni2023mattergen}: towards this aim, data quality remains key in determining reliability of AI-models \cite{himanen2019data}. 

Clearly, the quality of data is a multifaceted issue, as it is linked to disparate aspects in data generation including the accuracy by which materials properties are either measured \cite{germer2014spectrophotometry} or computed by simulations \cite{ohno2018computational}, the state of knowledge and/or ability to control operating parameters during experiments \cite{byl2012experimental}, the different adopted protocols and metrological approaches \cite{beauchamp2020metrological, garbe2020critical} etc.
In this work, we focus on a special aspect of data quality, potentially leading to a detrimental impact on the effectiveness of AI-based models to serve as platforms for materials discovery, namely biased sample selection.
Indeed, it is fair to expect that materials published in the literature and subsequently included within specialized databases were not randomly picked and tested over the years. 
Conversely, such materials, even those not exhibiting high performance according to a certain target property, have likely been carefully selected mostly based on the intuition and experience from field scientists.
Such prior experience can be regarded as a {\it latent knowledge} possessed by experts, who use it (either consciously or not) to make their choice before any experiments or studies is even initiated.
In other words, we expect that experimentalists will act each time in such a way to maximize the chances that the tested material has high performance.
As a result, the selection of tested material is never a fully random process, manifesting in an uneven and non-homogeneous representation of the materials space within a given database, and turning out in an anthropogenic bias \cite{jia2019anthropogenic}.
%


The issue is indeed well-known \cite{fujinuma2022big} and, along with the ongoing goal of extrapolative ML \cite{back2024accelerated, shimakawa2024extrapolative}, has been discussed in pertinent research articles within the field.
In particular, Kumagai \emph{et al.}~\cite{kumagai2022effects} argue that a direct effect of varying biases across different materials databases is the different elements distributions in terms of average atomic mass and average atomic electronegativity within a compound.
Interestingly, we show here that considering only these distributions is not enough to address data bias.
On the SuperCon database \cite{supercon_new}, the problem was also hypothesized and shortly discussed by Stanev \emph{et al.}~\cite{stanev2018machine}, who indeed included $\sim300$ materials found by the Hosono group not to display superconductivity \cite{hosono2015exploration} for mitigating the bias.
Nonetheless, such additional non-superconductors are anyway biased by human intuition towards the presence of superconductivity.

%
%
%

Sutton \emph{et al.}~\cite{sutton2020identifying} propose a methodology based on the identification of the domain of applicability of a trained model by means of a subgroup discovery (SGD) \cite{atzmueller2015subgroup} tool. 
Specifically, such an approach aims to determine the specific boundaries of each easily interpretable features (even starting from more  complex representations), thus identifying the domain of applicability as a \emph{hyperrectangle} in the feature space.
%
%
However, this methodology works on explicitly extracted features. 
As a consequence, this might be extended to Graph Neural Networks in a message passing fashion only converting such graph to a feature array by means e.g., a Variational Autoencoder, and subsequently applying such procedure over the learned latent space.


Li \emph{et al.}~\cite{li2023critical} and De Breuck \emph{et al.}~\cite{de2021robust} propose to employ methodologies based on unsupervised learning to identify the cluster of applicability of an already trained model.
%
%
%
However, the predictive power of clustering techniques is generally weaker than supervised methods, since clustering is designed for pattern discovery rather than making predictions on new, unseen data \cite{rokach2005clustering}.

Furthermore, De Breuck \emph{et al.}~\cite{de2021robust} also introduce an ensemble approach based on the Material Optimal Descriptor Network (MODNet) \cite{de2021materials} architecture, which provides both the predicted value of the property of interest and its associated uncertainty. 
They argue that for out-of-the-box samples, higher uncertainty indicates lower reliability in the predictions. 

All such reported methodologies are constructed exclusively over the same domain of the trained model. 
%
%
On the contrary, we propose here an alternative potential approach to detect, quantify and circumvent data bias in the field of materials discovery based also on out-of-domain samples.

Specifically, it consists of a set of binary classifier-based filters,  trained over samples from the specialized database (class 1) and random subsets from a \emph{less biased} general-purpose database like MaterialsProject (class 0), designed to rule out those out-of-the-box materials for which the ML prediction would not be reliable.
It is noteworthy that other studies in the literature \cite{stanev2018machine, gibson2024accelerating} utilize a pipeline encompassing classification. 
However, such classification was designed to exclude materials that are predicted to have a property of interest (i.e., superconducting critical temperature in the cited instances) below a certain threshold. 
Since such classifiers were trained, validated, and tested always over the same materials used in the regression model, they are not suitable to detect the data bias discussed above.


%
Herein, we concentrate on two relevant representative case studies, namely superconducting and thermoelectric materials.
As a result, in both case studies our methodology effectively rules out those samples for which regression predictions would not be reliable, proving that our approach helps to avoid cross-domain data bias in the field of materials discovery.

\section{Methods}

\subsection{Cross-domain data bias: assessing and circumventing}
The predominant approach in training and assessing ML algorithms involves the random partitioning of a single data source into training and test sets.
While this methodology is conventional, it overlooks a significant issue, i.e., the susceptibility to dataset bias \cite{torralba2011unbiased}.
%
%
For instance, a model trained exclusively on metals for the prediction of a property of interest will presumably lack the capacity to predict the same property for non-metallic out-of-the-box samples.

Let two random variables be defined, i.e., the signal $S$ and the bias $B$, serving as indicators in the identification process of an input as a specific target variable $Y$ \cite{bahng2020learning}. 
However, while the signal $S$ is essential for inferring the target $Y$, the bias $B$ is not (as far as the physical phenomenon is concerned). 
Therefore, taking advantage of the nomenclature introduced in ref.~\cite{bahng2020learning}, in this study we consider two learning scenarios depending on the relationship between the training distribution $p(S^{\mathrm{tr}}, Y^{\mathrm{tr}}, B^{\mathrm{tr}})$ and the out-of-the-box distribution $p(S^{\mathrm{out}}, Y^{\mathrm{out}}, B^{\mathrm{out}})$. 
In the former, i.e., the ``in-distribution" scenario, $p(S^{\mathrm{tr}}, Y^{\mathrm{tr}}, B^{\mathrm{tr}}) = p(S^{\mathrm{out}}, Y^{\mathrm{out}}, B^{\mathrm{out}})$. 
This is the typical case in which the out-of-the-box samples come from the same data source of the training samples. 

\begin{figure}
    \centering
    \includegraphics[width = 0.8\textwidth]{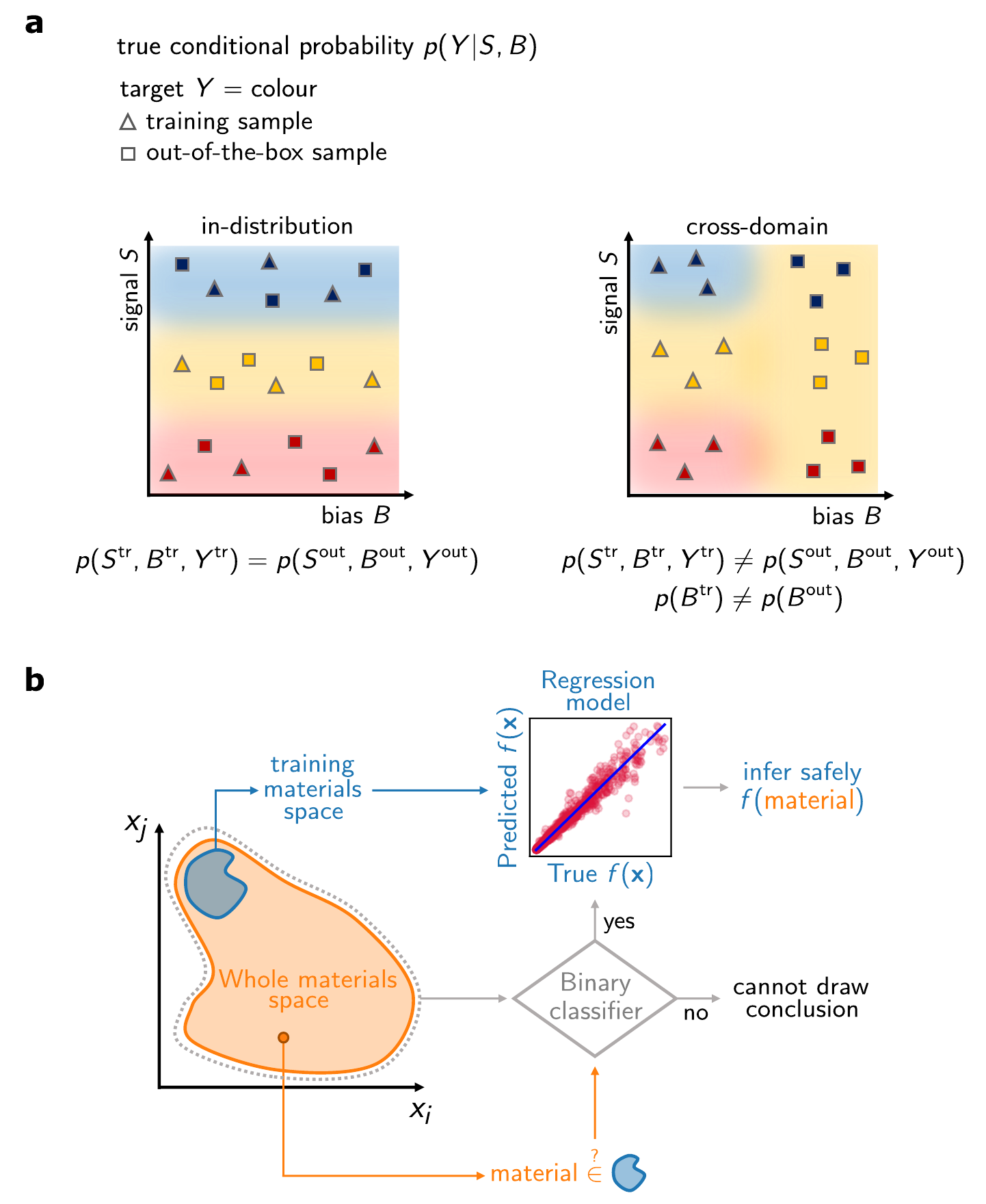}
    \caption{Overview of the main bias types and of the proposed methodology to circumvent it in materials discovery. \textbf{a} Sketches of in-distribution and cross-domain data-biases (adapted from ref.~\cite{bahng2020learning}). In the former, training samples and out-of-the-box samples share the same distributions in terms of the signal $S$, the bias $B$ and the target colour $Y$ (e.g., training samples and out-of-the-box samples come from the same source of data); in the latter, bias distributions are different (e.g., training samples and out-of-the-box samples come from the different sources of data). \textbf{b} Proposed methodology to detect and circumvent bias in materials discovery: a regression model is trained for the prediction of a target property $f(\mathbf{x})$; when predicting $f(\mathbf{x})$ for an out-of-the-box material, such prediction is reliable only if the material belongs to the same training materials space, otherwise no conclusion can be drawn; this is assessed by means of a binary classifier trained on the whole materials space.}
    \label{fig:bias}
\end{figure}

In the latter, i.e., the ``cross-domain" scenario, $p(S^{\mathrm{tr}}, Y^{\mathrm{tr}}, B^{\mathrm{tr}}) \neq p(S^{\mathrm{out}}, Y^{\mathrm{out}}, B^{\mathrm{out}})$, and also $p(B^{\mathrm{tr}}) \neq p(B^{\mathrm{out}})$; see Fig.~\ref{fig:bias}a, adapted from  ref.~\cite{bahng2020learning}. 
%
For instance, training data consist of materials with ($Y^{\mathrm{tr}}=\textrm{superconductor}$, $B^{\mathrm{tr}}=\textrm{metal}$) and ($Y^{\mathrm{tr}}=\textrm{non superconductor}$, $B^{\mathrm{tr}}=\textrm{metal}$), while out-of-the-box samples contain ($Y^{\mathrm{out}}=\textrm{superconductor}$, $B^{\mathrm{out}}=\textrm{non metal}$) and ($Y^{\mathrm{out}}=\textrm{non superconductor}$, $B^{\mathrm{out}}=\textrm{non metal}$).

Within this framework, the performance of the model on the out-of-the-box domain depends both on the performance observed in the training domain and on the degree of similarity existing between the two domains \cite{ben2006analysis}.
To take into account such similarity in the Materials Science context, we propose to employ a binary classifier, labeling with class 1 the samples in the training database (i.e., falling in the training domain) and with class 0 the samples, not included in the training database, from a broader-purpose database ideally covering the entirety of the materials space.

If the classifier is skilled, i.e., is able to correctly discriminate such samples over the two classes, than the training materials space is a ``localized" subset of the general materials space.
Thus, a ML model, even exhibiting high performances for the prediction of a property of interest on the training domain, cannot be used safely to predict the same property over all the out-of-the-box samples. 
In such a context, the aforementioned binary classifier can be used to filter out samples not belonging to the training domain, inferring the property of interest only for the out-of-the-box samples within the training domain (see Fig.~\ref{fig:bias}b).
Also, to avoid unfairness in the protocol, it should be ensured that such out-of-the-box samples are not already included in the class 0 labeled samples of the binary classifier training set; otherwise, those will likely be filtered out by the classifier itself.

\subsection{Validity checks}
To check the validity of such a methodology we propose a proof in two steps: (i) demonstrating the relationship between classifier performances and regressor reliability; (ii) proving that the MaterialsProject database is a suitable choice for a \emph{less} biased database, towards the accurate prediction of the property of interest.

\begin{figure}
    \centering \includegraphics[width=0.8\textwidth]{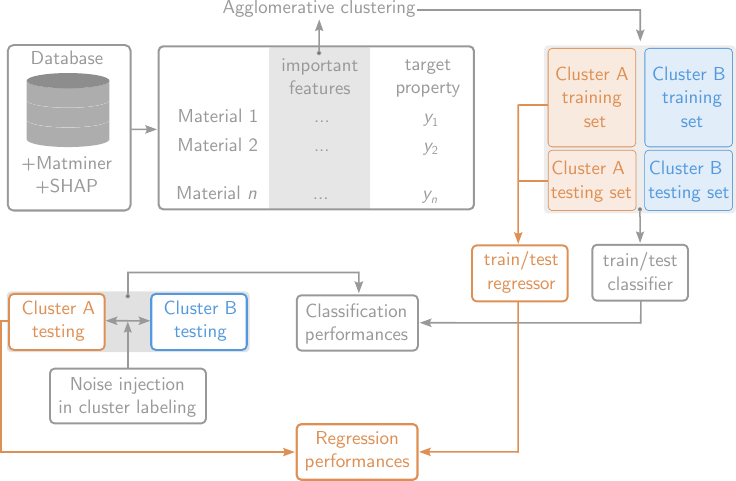}
    \caption{Protocol for validating the relationship between regressor and classifier performances (first validity check). 
    The specialized dataset comes with the most important Matminer composition-based features according to a SHAP ranking performed over the corresponding regression model. 
    A partition of such a dataset in two clusters, namely A and B, is obtained with the agglomerative clustering algorithm. 
    Furthermore, the entire dataset is randomly split in an 80/20 partition for training/testing of (i) a classifier for discriminating the two clusters and of (ii) a regressor for the target property of interest $y$ prediction only over cluster A. 
    The trained classifier is employed to discriminate the cluster A/B over the testing set, as the regressor is employed to predict $y$ over testing samples labeled as cluster A, with noise being progressively injected in cluster A/B labeling.
    In this way it is possible to assess the relationship between classification and regression performances. 
    }
    \label{fig:bias_confirm_1}
\end{figure}

The former validation, as depicted in Fig.~\ref{fig:bias_confirm_1}, can be achieved by properly clustering the specialized dataset.
Specifically, once a regression model is trained over the specialized dataset towards the prediction of a property of interest, a SHapley Additive exPlanations (SHAP)-based analysis can be conducted to get insight of the most important features. 
As a consequence, the SHAP analysis is conducted under the TreeSHAP routine, able to compute the Shapley values without approximations \cite{lundberg2017unified, lundberg2020local}.
Specifically, the importance of a descriptor is determined by comparing the output of a model that was trained with that particular feature to the output of a model trained without that feature (see refs. \cite{lundberg2017unified, lundberg2020local} for further details). 
Thus, we compute such coefficients of importance over the testing set.
Based on those important features only, we partition the specialized dataset into two clusters A and B by means of the Agglomerative Clustering algorithm \cite{scikit-learn}. 
Furthermore, to identify which key features to be included for Agglomerative clustering, we employ a procedure based on the Silhouette score, measuring how close each data point is to others in its own cluster and how far away the data point is from points in other clusters \cite{scikit-learn}. 
%
Further details in this respect are given in Supplementary Note 2.
%
%
At the same time, we split the entire dataset into a training set (80\%) and a testing set (20\%). We thus train a classifier with all the important features for discriminating the two identified clusters.
Additionally, we train a further regressor with all the important features intended to predict the property of interest only for testing set materials belonging to cluster A, with deliberately limiting its training set to only those training set materials coming from cluster A.
The idea is to progressively introduce random noise to the cluster labels of the testing set materials.
If the resulting gradual decrease in classifier performance during testing is associated with a decrease in regressor performance to predict the property of interest specifically for the testing set materials labeled as belonging to cluster A, it indicates a relationship between the performances of the classifier and the regressor.

\begin{figure}
    \centering \includegraphics[width=0.8\textwidth]{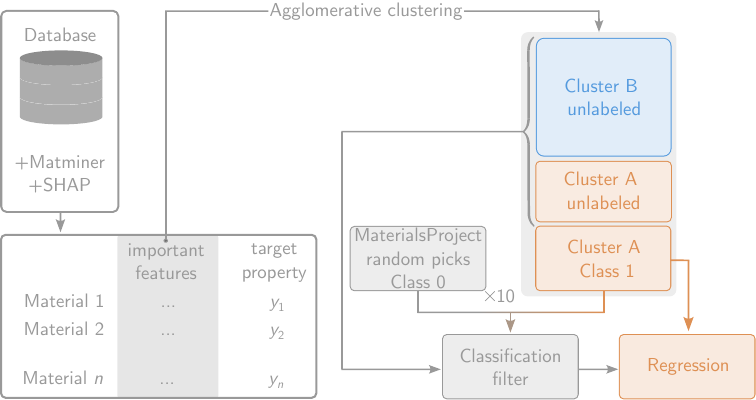}
    \caption{Overview of the protocol for validating the relationship between the classifier filtration stringency and the regression performances, along with the choice of MaterialsProject as \emph{less biased} database (second validity check). 
    The specialized dataset comes with the most important Matminer composition-based features according to a SHAP ranking performed over the corresponding regression model. 
    A partition of such a dataset in two clusters, namely A and B, is obtained with the agglomerative clustering algorithm as reported in Supplementary Note 2. 
    Half of cluster A is utilized to train/test a regressor for the prediction of the property of interest $y$. 
    The same half is utilized as class 1 across a set of 10 classifiers, with class 0 being represented by 10 different random subsets of the MaterialsProject database, each with the same cardinality of class 1. 
    The regression model is employed to predict the $y$ of those materials belonging to the second half of cluster A and to cluster B passing the classifier filtration, i.e., showing an average probability greater than a set threshold to be classified as class 1.}
    \label{fig:bias_confirm_2}
\end{figure}

The latter validation, as depicted in Fig.~\ref{fig:bias_confirm_2}, can be achieved with an analogous clustering of the specialized dataset.
However, in this instance we utilize only half of cluster A to train and test a regression model for the prediction of the property of interest.
Additionally, we utilize the same half of cluster A as class 1 across a set of 10 classifiers, while class 0 is represented by 10 different random subsets, each containing the same number of class 1 samples; such class 0 samples are MaterialsProject compositions not included in the specialized dataset.
If those classifiers prove to be highly skilled, we apply the regressor to the materials -- from the second half of cluster A as well as the entire cluster B -- which exhibit an average probability exceeding a specified threshold for being classified as class 1 (i.e., belonging to the SuperCon materials space) by the set of 10 classifiers. 
Specifically, we gradually increase this classification probability threshold from 0 (minimum stringency of the classifier-based filter) to 1 (maximum stringency of the classifier-based filter).
Thus, if the regression performances increase by increasing such stringency threshold, this proves that the MaterialsProject database is a proper choice as source of \emph{less} biased samples and, more importantly, the proposed methodology is able to override cross-domain bias.

\subsection{Supervised ML models}

The classification models used in the detection and quantification of data bias, along with those used to compare with the hypothesis by Kumagai \emph{et al.}~\cite{kumagai2022effects} consist of Extra Trees Classifier (ETC)-based pipelines \cite{scikit-learn}, trained over the 80\% of the respective datasets and tested over the remaining 20\% with hyperparameter tuning in stratified 5-fold cross validation.
Such pipeline encompasses linear correlation analysis for feature reduction, variance analysis of descriptors, correlation analysis with the property of interest, and ML with hyperparameter tuning in 5-fold cross-validation. The hyperparameter space explored in such cross-validation is given in Supplementary Note 1.
Conversely, all the classification models used in the validity check procedure consist of Extra Trees Classifier (ETC)-based models \cite{scikit-learn}, trained over the 80\% of the respective datasets and tested over the remaining 20\% with default hyperparameters.
Specifically, the Scikit-learn Python package \cite{scikit-learn} provides the capability to not only predict the class but also estimate class probabilities. 
The predicted class is automatically determined as the one with the highest probability. 
Consequently, when focusing only on the probabilities of class 1, signifying the prediction of the material to be in SuperCon, we adjusted the discrimination threshold from 0 (predicting all materials as class 1) to 1 (predicting all materials as class 0). 
For each threshold value, a distinct confusion matrix was constructed, resulting in varying counts of true positives (TP), false negatives (FN), false positives (FP), and true negatives (TN). 
These matrices were used to compute the true positive rate (TPR) and the false positive rate (FPR), where $\rm{TPR} = TP / (TP + FN)$ and $\rm{FPR} = FP / (FP + TN)$. 
The classifier performance can be assessed using Area Under Curve (AUC) of the Receiver Operating Characteristic (ROC), where larger AUCs indicate better performance. 

The regression models utilized for assessing the most important features underlying a property of interest consist of Extra Trees Regressor (ETR)-based pipelines \cite{scikit-learn}, with hyperparameter tuning in 5-fold cross validation, trained/validated over the 80\% of the respective datasets and tested over the remaining 20\%.
Such pipeline encompasses linear correlation analysis for feature reduction, variance analysis of descriptors, correlation analysis with the property of interest, and ML with hyperparameter tuning in 5-fold cross-validation. The hyperparameter space explored in such cross-validation is given in Supplementary Note 1. 
The remaining regression models -- used in the validity check procedure -- consist of ETR-based models, trained over the 80\% of the respective datasets and tested over the remaining 20\% with default hyperparameters.

\section{Results}

\subsection{Superconducting materials}

Superconductors exhibit no electrical resistivity when cooled below a superconducting critical temperature $T_{\rm{c}}$ \cite{hirsch2015superconducting}. 
Owing to this inherent characteristic, those compounds have captured interest across diverse fields with multiple applications, like (and not limited to) Superconducting Magnetic Energy Storage (SMES) systems, enabling energy storage through a direct current (DC) passing through a superconducting coil \cite{johnson2019selecting}; superconducting electromagnets, finding applications in fusion reactors like tokamak \cite{yuanxi2006first}, Magnetic Resonance Imaging (MRI) \cite{aarnink2012magnetic, hall1991use}, Nuclear Magnetic Resonance (NMR) machines \cite{asayama1996nmr, rigamonti1998basic},  particle accelerators \cite{rossi2012superconducting}.
Therefore, the identification of novel superconductors in the foreseeable future is greatly sought after and could wield a significant influence on sectors including the energy industry.

\subsubsection{SuperCon and featurization}
Many works in the literature \cite{stanev2018machine, konno2021deep, le2020critical, roter2020predicting, roter2022clustering, trezza2023leveraging} have been possible thanks to on a convenient source of data, namely the SuperCon database \cite{supercon_new}, which collects the values of experimentally measured critical temperatures $T_{\rm{c}}$ of materials whose superconductivity has been tested from a vast body of scientific literature.
Indeed, to the best of our knowledge, SuperCon can be considered as the most comprehensive database in the field, and this alone can explain the reason why the idea of using it for training AI based models for material discovery appears so tempting. 
%
%
Specifically, it gathers materials of both inorganic (classified as ``Oxide and Metallic") and organic origin (classified as ``Organic").
We considered only the entire subset of inorganic compounds, consisting of $\sim33,000$ entries, of which $\sim7,000$ have no $T_{\rm{c}}$; we dropped those latter compounds.
Also, we dropped all materials whose formulae contain symbols like `-', `+', `,', strings like `X', `Z', `z' when not included in meaningful elements symbols (e.g., `Zn'), and  \ch{Pb2CAg2O6} at $323\, \si{\kelvin}$ as it is wrong (\ch{LaH10} at $250\, \si{\kelvin}$ represents the superconductor with the highest known $T_{\mathrm{c}}$ \cite{drozdov2019superconductivity}). 
After normalizing the formulae stoichiometry, if the same compound was reported with multiple $T_{\rm{c}}$ values, we retained the average critical temperature only for those materials exhibiting a Relative Standard Deviation (RSD) of less than 20\% across those occurrences. This filtering process led to reduce the total number of compounds in the dataset to 12,804.
We stress that, in addition to the $T_{\rm{c}}$ values and some partial information about the pressure, the SuperCon database provides only the chemical composition of the corresponding compounds, of which it is thus not possible to unambiguously infer the structures.
For this reason, following a quite popular approach in the literature \cite{stanev2018machine} we utilized Matminer \cite{ward2018matminer} which enables to associate 145 composition-based descriptors to the normalized brute formula of each compound. Specifically, as outlined by Ward \emph{et al.}~\cite{ward2016general}, these descriptors encompass stoichiometric characteristics, statistics on elemental properties, attributes related to electronic structure, and attributes specific to ionic compounds (see also Supplementary Note 1 for further details).

\subsubsection{Cross-domain bias detection and quantification}

In order to detect and quantify data bias, and also testing the hypothesis by Kumagai \emph{et al.}~\cite{kumagai2022effects}, we create 4 datasets, namely A$^{\prime}$, B$^{\prime}$, C$^{\prime}$, D$^{\prime}$ suitable for classification tasks.

Specifically, dataset A$^{\prime}$ contains the 12,804 featurized formulae from the SuperCon database labeled as class 1, along with 12,804 featurized randomly picked formulae from MaterialsProject (not included in SuperCon) labeled with class 0. 
The ETC-based pipeline trained over dataset A$^{\prime}$ is highly skilled in testing, showing an $\mathrm{AUC}\approx0.99$, thus suggesting that the SuperCon database is effectively localized with respect to the MaterialsProject database.
Indeed, we create dataset B$^{\prime}$, which contains the very same formulae, but with random labeling of the classes 0 and 1.
As expected, in this case the classifier is non-skilled, with $\mathrm{AUC}\approx0.5$.

Also, beyond these two extreme cases and following upon the claim by Kumagai \emph{et al.}, if materials from the specialized and the general database present the same distribution in terms of two descriptors -- namely atomic weight and electronegativity -- it should not be possible to set up a highly skilled classifier as in the test above.
\begin{figure}
    \centering
    \includegraphics[width=0.8\linewidth]{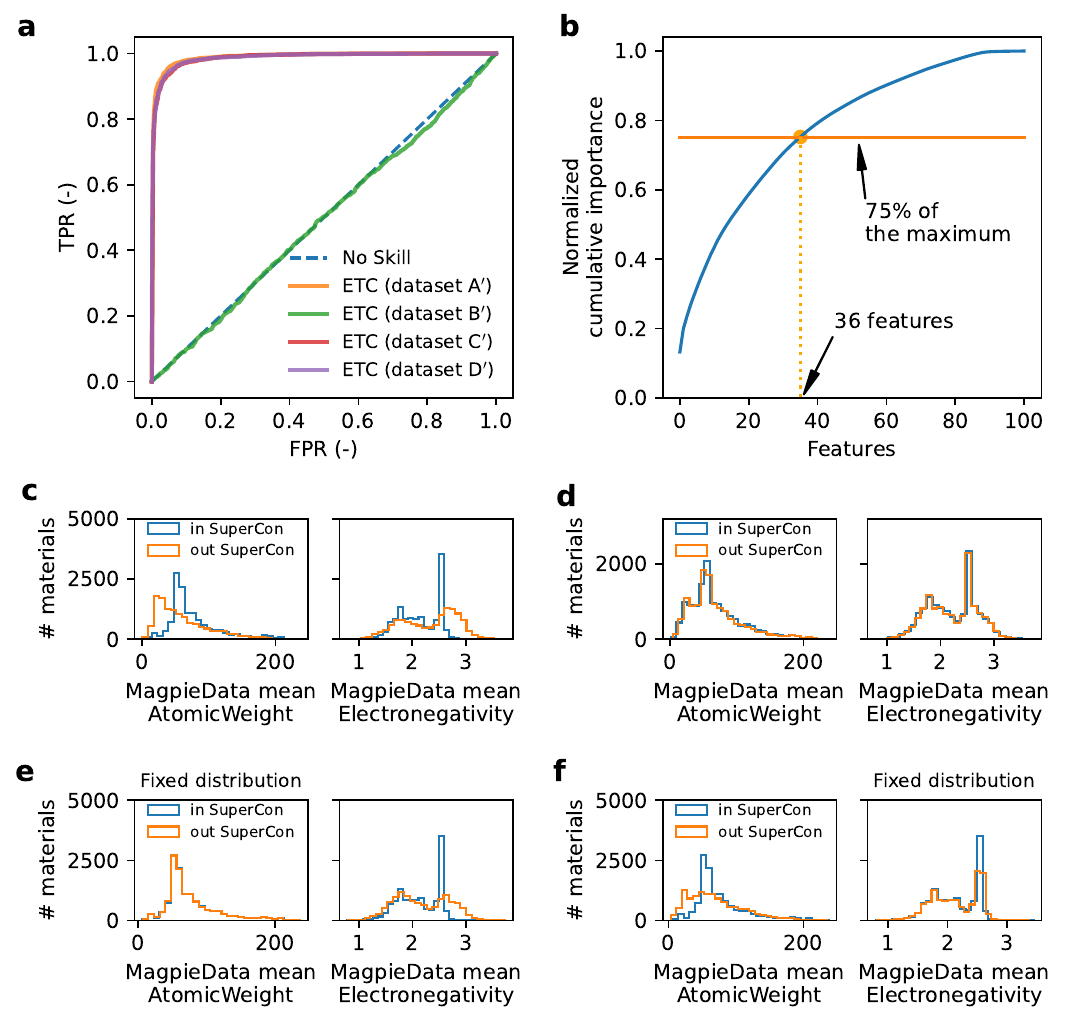}
    \caption{ETC-based pipelines results and datasets compositions. 
    $\mathbf{a}$ ROC curves for classification models over datasets A$^{\prime}$, B$^{\prime}$, C$^{\prime}$, D$^{\prime}$, as in the main text, together with No Skill classifier. 
    $\mathbf{b}$ Normalized cumulative curve for the coefficients of importance of the ETC-based pipeline on dataset A$^{\prime}$.
    $\mathbf{c}$ Distributions over dataset A$^{\prime}$ for the features ``MagpieData mean AtomicWeight" and ``MagpieData mean Electronegativity" of materials in SuperCon (blue) and out SuperCon (orange).
    $\mathbf{d}$ Distributions over dataset B$^{\prime}$ for the features ``MagpieData mean AtomicWeight" and ``MagpieData mean Electronegativity" of materials in SuperCon (blue) and out SuperCon (orange).
    $\mathbf{e}$ Distributions over dataset C$^{\prime}$ for the features ``MagpieData mean AtomicWeight" and ``MagpieData mean Electronegativity" of materials in SuperCon (blue) and out SuperCon (orange). 
    $\mathbf{f}$ Distributions over dataset D$^{\prime}$ for the features ``MagpieData mean AtomicWeight" and ``MagpieData mean Electronegativity" of materials in SuperCon (blue) and out SuperCon (orange).}
    \label{fig:ETC_supercon_bias}
\end{figure}
To this end, we construct two further datasets, namely C$^{\prime}$ and D$^{\prime}$. 
Specifically, dataset  C$^{\prime}$ is obtained by adding, to the same 12,804 materials from SuperCon, 12,804 random materials from MaterialsProject following the same probability distribution of the average atomic mass (feature ``MagpieData mean AtomicWeight") of the SuperCon database.
Analogously, dataset D$^{\prime}$ is obtained by adding, to the same 12,804 materials from SuperCon, 12,804 random materials from MaterialsProject following the same probability distribution of the average electronegativity (feature ``MagpieData mean Electronegativity") of the SuperCon database.
Still in those two further cases the classifiers are highly skilled in testing, with $\mathrm{AUC}\approx0.99$.

The ROC curves of those four classifiers are showcased in Fig.~\ref{fig:ETC_supercon_bias}a. 
In principle, we recognize that a classifier may appear highly skilled as it learns seemingly simple or trivial patterns, such as shortcuts relying on a limited number of features \cite{geirhos2020shortcut}.
To rule out such possibility and gain insights into the model behaviour, we utilize SHAP for interpretation over the model trained/tested on dataset A$^{\prime}$. 
The latter analysis reveals that a substantial amount (i.e., 36) of features contribute to 75\% of the model's output importance (see Fig.~\ref{fig:ETC_supercon_bias}b), thus suggesting that the trained classifier is non-trivial.
Also, Figs.~\ref{fig:ETC_supercon_bias}c, d, e, f report the distributions of materials labeled as ``in SuperCon" or ``out SuperCon" in terms of the two features ``MagpieData mean AtomicWeight" and ``MagpieData mean Electronegativity" over the four datasets A$^{\prime}$, B$^{\prime}$, C$^{\prime}$, D$^{\prime}$ respectively. 
Specifically, the distributions for dataset A$^{\prime}$ (Fig.~\ref{fig:ETC_supercon_bias}c) show a clear distinction between in and out SuperCon samples, whereas the distributions for dataset B$^{\prime}$ (Fig.~\ref{fig:ETC_supercon_bias}d) are nearly identical, as expected, due to the random assignment of labels.
For dataset C$^{\prime}$ (Fig.~\ref{fig:ETC_supercon_bias}e), the distribution of atomic weight is intentionally kept identical for both in and out SuperCon instances, leading to the distribution of electronegativity being approximately similar between the two groups to some extent.
The same applies, with inverted features, to dataset D$^{\prime}$ (Fig.~\ref{fig:ETC_supercon_bias}f).
%
One could argue that this analysis might introduce potential unfairness in classifier performance due to material classes in SuperCon, such as doped compounds, which are generally absent from the MaterialsProject. 
To address this, we conducted an additional analysis considering only the materials from the SuperCon database that are also present in the MaterialsProject, and the results remain consistent (see Supplementary Note 3).
Importantly, here we suggest that the AUC can be regarded as a measure of the SuperCon bias relative to the MaterialsProject bias and that the higher the AUC, the higher the bias.

\subsubsection{Features importances}

Here, we aim to identify the most important features for predicting the $T_{\mathrm{c}}$, which will be effectively utilized within the two validity checks in the next subsection.

\begin{figure}
    \centering
    \includegraphics[width = 0.8\textwidth]{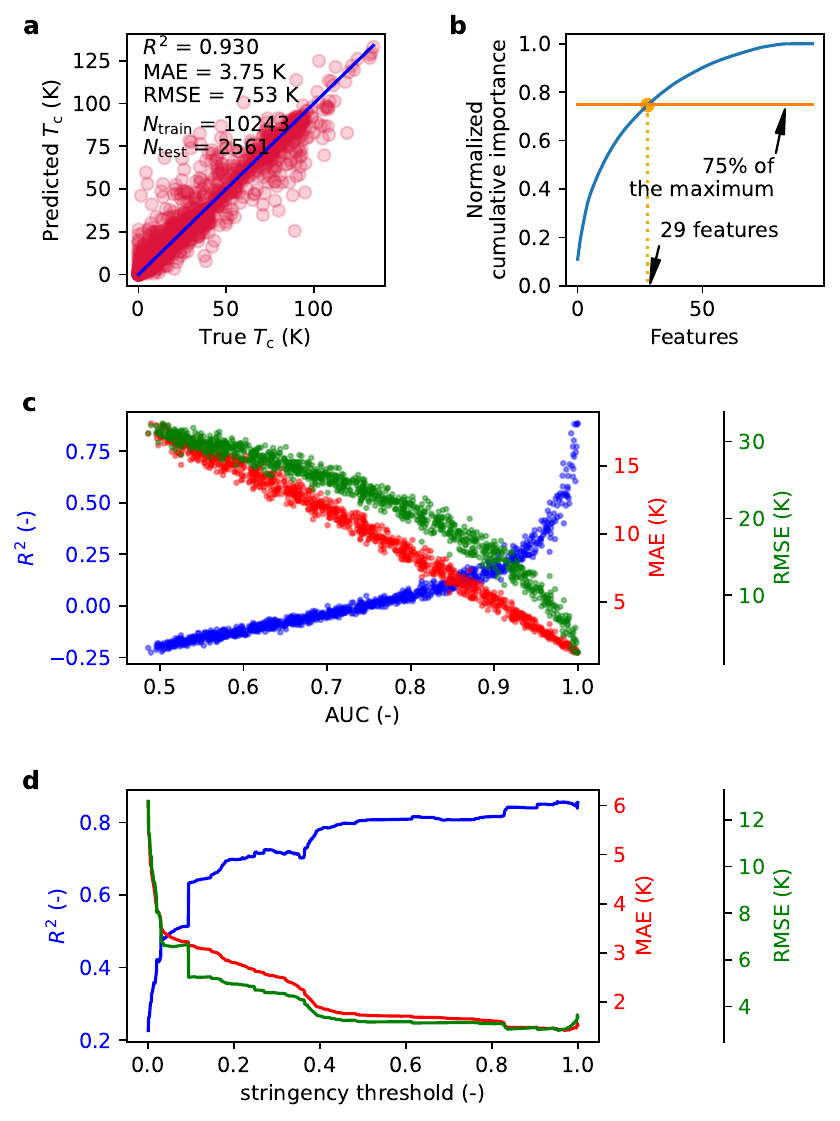}
    \caption{Results of the validity checks for the superconductors case.
    \textbf{a} Predictions of the ETR-based pipeline over the SuperCon testing set along with corresponding performances shown in terms of coefficient of determination $R^2$, Mean Absolute Error (MAE), Root Mean Squared Error (RMSE), with the size of training and testing sets $N_{\textrm{train}}$ and $N_{\textrm{test}}$, respectively. 
    \textbf{b} Corresponding normalized cumulative curve for the SHAP-based coefficients of importance. 
    \textbf{c} Results of the first validity check (as detailed in the main text) in terms of the performances ($R^2$, MAE, RMSE) for a default-hyperparameters ETR regressor trained with the 29 most important features as from subplot \textbf{b} over cluster A only, compared with respect to the AUC of an ETC classifier with default hyperparameters trained for discriminating the two clusters of the SuperCon database, for 1000 distinct noise injections percentages in cluster labels.
    \textbf{d} Results of the second validity check (as detailed in the main text) in terms of the performances ($R^2$, MAE, RMSE) for the same ETR regressor detailed in subplot \textbf{c} compared with respect to 1000 distinct stringency threshold values, 
    each computed as the threshold above which the average probability over 10 ETC-based classifiers trained with the same 29 features as in subplot \textbf{b} with default hyperparameters has to be set for materials in the testing set to be classified as in the same ETR training materials space.
    }
    \label{fig:bias_confirm_SuperCon}
\end{figure}

As part of the data pre-processing routines, the pipeline used in this work, encompassing linear correlation analysis for feature reduction, variance analysis of descriptors, correlation analysis with $T_{\rm{c}}$, and ML with hyperparameter tuning in 5-fold cross-validation (refer to Supplementary Note 2 for details), drops 50 of the 145 features.
This underscores that numerous descriptors initially chosen have a limited impact on the designated target property (here the critical temperature).
As usual, we use $80\%$ of the dataset for training/cross-validation and the remaining $20\%$ for testing; as depicted in Fig.~\ref{fig:bias_confirm_SuperCon}a, the Extra Trees Regressor (ETR)-based pipeline is highly predictive, achieving a coefficient of determination $R^2=0.930$ over the testing samples, never encountered by the model during the training.

%

%

The SHAP-based analysis under the TreeSHAP routine, optimal for ensemble decision trees-based ML models like ETR \cite{lundberg2017unified, lundberg2020local}, is able to identify the most relevant descriptors for the trained model among the effectively employed 95 features retained after the preprocessing steps above.
%
%
It follows that 29 features account for the 75\% of the cumulative curve over the normalized coefficients of importance, as shown in Fig.~\ref{fig:bias_confirm_SuperCon}b.

\subsubsection{Validity checks}
Once the most important features are correctly identified, we proceed with the two validity checks already described in the Methods section.
%

%


We first clusterize the 12,804 SuperCon materials in two clusters A and B by means of an agglomerative clustering algorithm according to the procedure described in Supplementary Note 2, leading to cluster A with 4057 materials, and to cluster B with 8747 materials.

We therefore follow the procedure outlined above in the Methods section to detect any potential correlation between the classification and regression performances. 
We recall this -- adopting all the 29 important features as identified by SHAP -- involves training and testing an ETR on cluster A alone, and training and testing an ETC on both clusters A and B, thus using the ETR to predict the $T_{\mathrm{c}}$ of labeled instances from cluster A only, while injecting noise into the cluster labels and lowering its testing AUC.
In this respect, Fig.~\ref{fig:bias_confirm_SuperCon}c confirms that the larger AUCs in testing, the better performances of the regressor in terms of $R^2$, MAE, RMSE. 
This shows a strong correlation between classification and regression performances: the more effectively the classifier can discriminate between clusters, the better the regressor will perform on instances within its respective training cluster.

However, this initial validation check does not consider the selection of MaterialsProject as a \emph{less} biased source of samples, nor it accounts for the choice of a stringency probability threshold above which an out-of-the-box material is classified as belonging to the same materials space as the regression training and testing sets.
To this end, we follow the procedure outlined above in the Methods section for the second validity check.
This involves using only half of cluster A to train and test a regression model for predicting the $T_{\mathrm{c}}$ and employing the same half of cluster A as class 1 across a set of 10 classifiers, while class 0 is represented by 10 different random subsets, each containing the same number of samples as class 1. 

Such class 0 samples are MaterialsProject compositions not included in the SuperCon database.
We apply the regressor to predict the $T_\mathrm{c}$ only for materials in the second half of cluster A and all of cluster B that have an average probability above a specified stringency threshold to be classified as belonging to class 1 and, equivalently, belonging to cluster A (i.e., the same regression training/testing cluster).
In this respect, Fig.~\ref{fig:bias_confirm_SuperCon}d shows that the larger the stringency threshold in testing, the better (albeit non monotonously) performances of the regressor in terms of $R^2$, MAE, RMSE. 
This confirms that our methodology is able to effectively rule out those out-of-the-box materials not belonging to the same materials space as the regression model training/testing data, and for which the regression prediction would not have been reliable.

\subsection{Thermoelectric materials}
On the other hand, thermoelectric materials exhibit a strong coupling between electrical and thermal transport phenomena, thus enabling the direct conversion of temperature gradients into electrical voltage and vice versa \cite{gayner2016recent}. 
As a consequence, thermoelectric-based devices can in principle harness various heat sources, like solar radiation and industrial waste heat and, as such, are crucial for developing sustainable and energy-efficient technologies.
The performance of a thermoelectric material is quantified by the dimensionless thermoelectric figure of merit, given by $zT = (S^2\sigma/\kappa)T$; here, $S$ represents the Seebeck coefficient, $\sigma$ the electrical conductivity, $\kappa$ the thermal conductivity, all of which depend on the temperature $T$.

\subsubsection{ESTM and featurization}

In this instance, we utilize the Experimentally Synthesized Thermoelectric Materials (ESTM) database \cite{na2022public} as data source, which provides the experimental $zT$ values for 872 materials as a function of $T$.

Still in this case only the composition is available and, as such, we employ the same 145 composition based features.
Additionally, because each material in ESTM is associated with multiple distinct temperature values -- which in general vary across the database -- we treat $T$ as the 146\textsuperscript{th} feature.
However, to avoid potential unfairness that could result from having the same material -- differing only in the temperature -- appear across different random splits in the training, validation, and testing sets we keep each material only in the instance with its minimum $T$ value.

\subsubsection{Cross-domain bias detection and quantification}

Analogously to the example above, to detect and quantify data bias, and also testing the hypothesis by Kumagai \emph{et al.}~\cite{kumagai2022effects}, we create 4 datasets, namely A$^{\prime\prime}$, B$^{\prime\prime}$, C$^{\prime\prime}$, D$^{\prime\prime}$ suitable for classification tasks.

Specifically, dataset A$^{\prime\prime}$ contains the 872 featurized formulae from the SuperCon database labeled as class 1, along with 872 featurized randomly picked formulae from MaterialsProject (not included in SuperCon) labeled with class 0. 
In particular, for those MaterialsProject samples, we set the $T$ -- namely the 146\textsuperscript{th} features -- at $329\, \si{\kelvin}$, which corresponds to the average temperature employed over the 872 ESTM materials.
The ETC-based pipeline trained over dataset A$^{\prime\prime}$ is highly skilled in testing, showing an $\mathrm{AUC}\approx0.99$, thus suggesting that the ESTM database is effectively localized with respect to the MaterialsProject database.
Indeed, we create dataset B$^{\prime\prime}$, which contains the very same formulae, but with random labeling of the classes 0 and 1.
As expected, in this case the classifier is non-skilled, with $\mathrm{AUC}\approx0.5$.

Also, as done above, we try to check the claim by Kumagai \emph{et al.}, thus constructing two further datasets, namely C$^{\prime\prime}$ and D$^{\prime\prime}$. 
Specifically, dataset  C$^{\prime\prime}$ is obtained by adding, to the same 872 materials from ESTM, 872 random materials from MaterialsProject following the same probability distribution of the average atomic mass (feature ``MagpieData mean AtomicWeight") of the SuperCon database.
Analogously, dataset D$^{\prime\prime}$ is obtained by adding, to the same 872 materials from ESTM, 872 random materials from MaterialsProject following the same probability distribution of the average electronegativity (feature ``MagpieData mean Electronegativity") of the SuperCon database.
Still in those two further cases the classifiers are highly skilled in testing, with $\mathrm{AUC}\approx0.99$.

\begin{figure}
    \centering
    \includegraphics[width=0.9\linewidth]{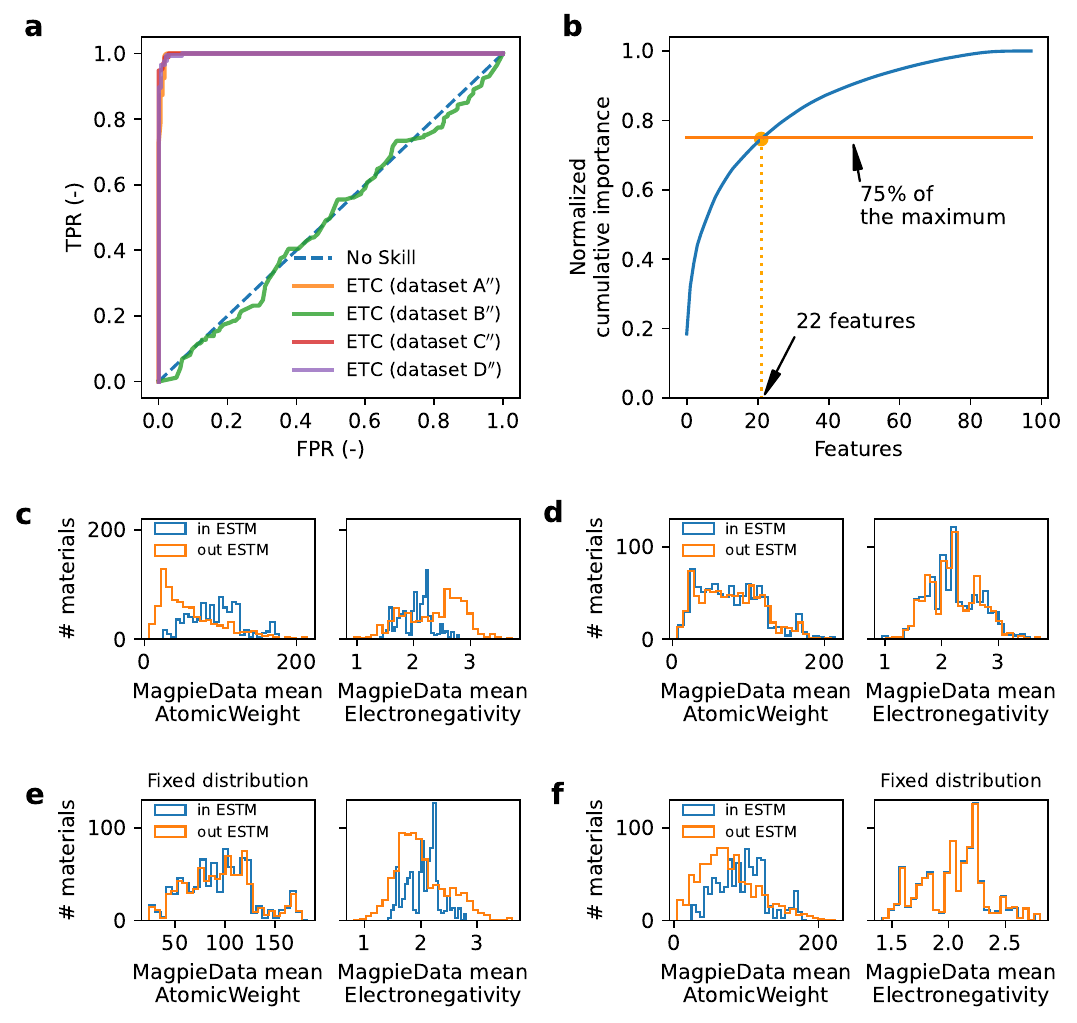}
    \caption{ETC-based pipelines results and datasets compositions. 
    $\mathbf{a}$ ROC curves for classification models over datasets A$^{\prime\prime}$, B$^{\prime\prime}$, C$^{\prime\prime}$, D$^{\prime\prime}$, as in the main text, together with No Skill classifier. 
    $\mathbf{b}$ Normalized cumulative curve for the coefficients of importance of the ETC-based pipeline on dataset A$^{\prime\prime}$.
    $\mathbf{c}$ Distributions over dataset A$^{\prime\prime}$ for the features ``MagpieData mean AtomicWeight" and ``MagpieData mean Electronegativity" of materials in ESTM (blue) and out ESTM (orange).
    $\mathbf{d}$ Distributions over dataset B$^{\prime\prime}$ for the features ``MagpieData mean AtomicWeight" and ``MagpieData mean Electronegativity" of materials in ESTM (blue) and out ESTM (orange).
    $\mathbf{e}$ Distributions over dataset C$^{\prime\prime}$ for the features ``MagpieData mean AtomicWeight" and ``MagpieData mean Electronegativity" of materials in ESTM (blue) and out ESTM (orange). 
    $\mathbf{f}$ Distributions over dataset D for the features ``MagpieData mean AtomicWeight" and ``MagpieData mean Electronegativity" of materials in ESTM (blue) and out ESTM (orange).}
    \label{fig:ETC_estm_bias}
\end{figure}

The ROC curves of those four classifiers are showcased in Fig.~\ref{fig:ETC_estm_bias}a. 
As done above, to gain insights into the model behaviour, we utilize SHAP for interpretation over the model trained/tested on dataset A$^{\prime\prime}$. 
Still in the case, the latter analysis reveals that a substantial amount (i.e., 22) of features contribute to 75\% of the model output importance (see Fig.~\ref{fig:ETC_supercon_bias}b), thus indicating that the trained classifier is non-trivial.
Also, Figs.~\ref{fig:ETC_estm_bias}c, d, e, f report the distributions of materials labeled as ``in ESTM" or ``out ESTM" in terms of the two features ``MagpieData mean AtomicWeight" and ``MagpieData mean Electronegativity" over the four datasets A$^{\prime\prime}$, B$^{\prime\prime}$, C$^{\prime\prime}$, D$^{\prime\prime}$ respectively. 
Specifically, the distributions for dataset A$^{\prime\prime}$ (Fig.~\ref{fig:ETC_estm_bias}c) show a clear distinction between in and out SuperCon samples, whereas the distributions for dataset B$^{\prime\prime}$ (Fig.~\ref{fig:ETC_supercon_bias}d) are nearly identical, as expected, due to the random assignment of labels.
For dataset C$^{\prime\prime}$ (Fig.~\ref{fig:ETC_estm_bias}e), the distribution of atomic weight is intentionally kept identical for both in and out SuperCon instances, leading to the distribution of electronegativity being approximately similar between the two groups to some extent.
The same applies, with inverted features, to dataset D$^{\prime\prime}$ (Fig.~\ref{fig:ETC_estm_bias}f).
%
Also in this case, we conducted an additional analysis considering only the materials from the ESTM database that are also present in the MaterialsProject, and the results remain consistent (see Supplementary Note 3).

As a consequence, still we would suggest that the AUC can be regarded as a measure of the ESTM bias relative to the MaterialsProject bias and that the higher the AUC, the higher the bias.

\subsubsection{Features importances}
Here, we aim to identify the most important features for predicting $zT$, which will be effectively utilized within the two validity checks in the next subsection.

As part of the data pre-processing routines, the pipeline used in this work, encompassing linear correlation analysis for feature reduction, variance analysis of descriptors, correlation analysis with $zT$, and ML with hyperparameter tuning in 5-fold cross-validation (refer to Supplementary Note 1 for details), drops 76 of the 146 features.
Still, this underscores that numerous descriptors initially chosen have a limited impact on the designated target property (here the critical temperature).
As usual, we use $80\%$ of the dataset for training/cross-validation and the remaining $20\%$ for testing; as depicted in Fig.~\ref{fig:bias_confirm_estm}a, the Extra Trees Regressor (ETR)-based pipeline is highly predictive, achieving a coefficient of determination $R^2=0.792$ over the testing samples, never encountered by the model during the training.
The SHAP-based analysis under the TreeSHAP routine, is able to identify the most relevant descriptors for the trained model among the effectively employed 70 features retained after the preprocessing steps above.
It follows that 32 features account for the 75\% of the cumulative curve over the normalized coefficients of importance, as shown in Fig.~\ref{fig:bias_confirm_estm}b.

\subsubsection{Validity checks}

As in the superconductor case, once the most important features are correctly identified, we proceed with the two validity checks already described in the Methods section.

First, we effectively clusterize the 872 SuperCon materials in two clusters A and B according to the methodology in Supplementary Note 2, which lead to cluster A with 454 materials, and to cluster B with 416 materials.

\begin{figure}
    \centering
    \includegraphics[width=0.7\linewidth]{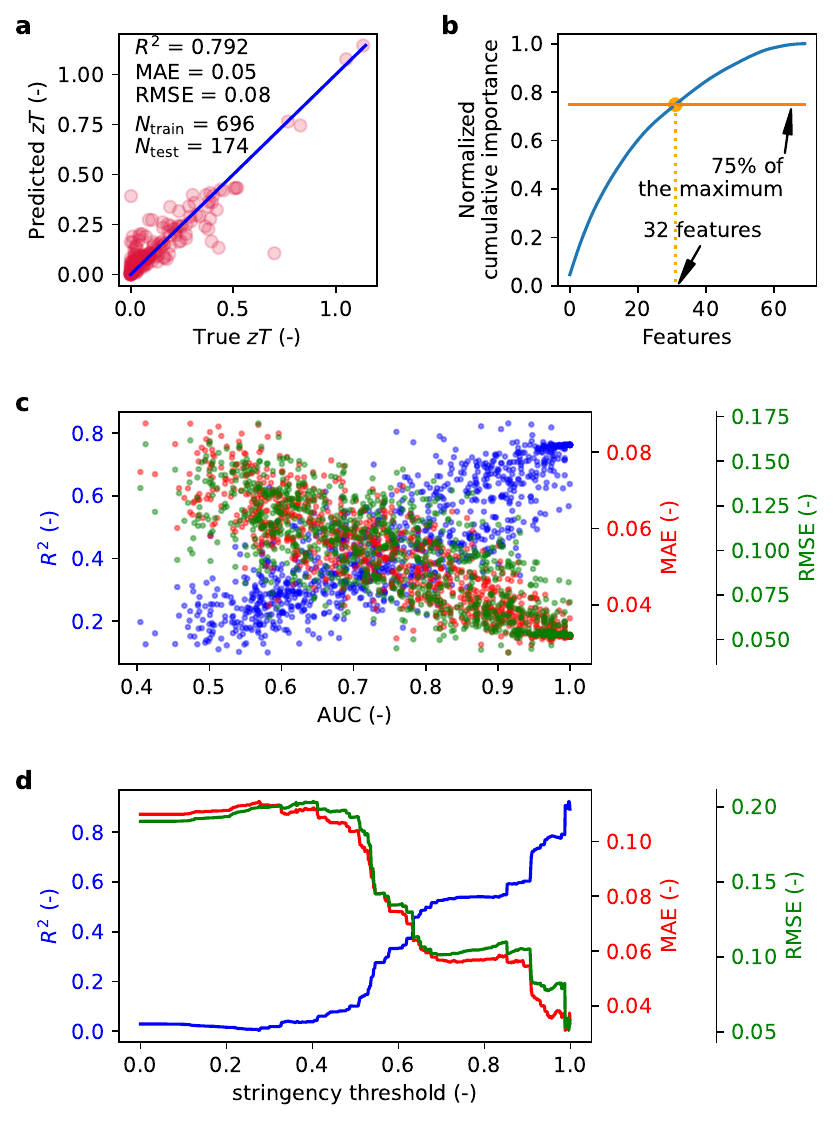}
    \caption{Results of the validity checks for the thermoelectrics case.
    \textbf{a} Predictions of the ETR-based pipeline over the ESTM testing set along with corresponding performances shown in terms of coefficient of determination $R^2$, Mean Absolute Error (MAE), Root Mean Squared Error (RMSE), with the size of training and testing sets $N_{\textrm{train}}$ and $N_{\textrm{test}}$, respectively. 
    \textbf{b} Corresponding normalized cumulative curve for the SHAP-based coefficients of importance. 
    \textbf{c} Results of the first validity check (as detailed in the main text) in terms of the performances ($R^2$, MAE, RMSE) for a default-hyperparameters ETR regressor trained with the 29 most important features as from subplot \textbf{b} over cluster A only, compared with respect to the AUC of an ETC classifier with default hyperparameters trained for discriminating the two clusters of the ESTM database, for 1000 distinct noise injections percentages in cluster labels.
    \textbf{d} Results of the second validity check (as detailed in the main text) in terms of the performances ($R^2$, MAE, RMSE) for the same ETR regressor detailed in subplot \textbf{c} compared with respect to 1000 distinct stringency threshold values, 
    each computed as the threshold above which the average probability over 10 ETC-based classifiers trained with the same 32 features as in subplot \textbf{b} with default hyperparameters has to be set for materials in the testing set to be classified as in the same ETR training materials space.
    }
    \label{fig:bias_confirm_estm}
\end{figure}

We therefore follow the procedure outlined above in the Methods section to detect any potential correlation between the classification and regression performances. 
We recall this involves training and testing an ETR on cluster A alone, and training and testing an ETC on both clusters A and B, thus using the ETR to predict the $zT$ of labeled instances from cluster A only, while injecting noise into the cluster labels and lowering its testing AUC.
In this respect, Fig.~\ref{fig:bias_confirm_estm}c confirms that the larger AUCs in testing, the better performances of the regressor in terms of $R^2$, MAE, RMSE. 
This shows a strong correlation between classification and regression performances: the more effectively the classifier can discriminate between clusters, the better the regressor will perform on instances within its respective training cluster.
Subsequently, we follow the procedure outlined above in the Methods section for the second validity check.
We recall this involves using only half of cluster A to train and test a regression model for predicting the $zT$ and employing the same half of cluster A as class 1 across a set of 10 classifiers, while class 0 is represented by 10 different random subsets, each containing the same number of samples as class 1. 
Such class 0 samples are MaterialsProject compositions not included in the SuperCon database.
We apply the regressor to predict the $zT$ only for materials in the second half of cluster A and all of cluster B that have an average probability above a specified stringency threshold to be classified as belonging to class 1 and, equivalently, belonging to cluster A (i.e., the same regression training/testing cluster).
In this respect, Fig.~\ref{fig:bias_confirm_estm}d shows that the larger the stringency threshold in testing, the better (albeit non monotonously) performances of the regressor in terms of $R^2$, MAE, RMSE. 
This still confirms that our methodology is able to effectively rule out those out-of-the-box materials not belonging to the same materials space as the regression model training/testing data, and for which the regression prediction would not have been reliable.

\section{Conclusion and discussion}
%
%
%
%

In this work we have proposed a methodology based on a set of classifiers for taking into account cross-domain bias of specialized datasets.
As such, before inferring the target property over samples belonging to a different source of data, we suggest that a check is made with a classifier based on both the specialized database of interest and a more general one as described above: only if such classification test is passed (i.e., the out-of-the-box sample belongs to the same space of the training database), than the regression model prediction may be considered reliable.
In this respect, the AUC may even serve as a quantitative measure of the data bias.

As mentioned earlier in the Introduction, alternative protocols in the literature are primarily based on unsupervised learning methods (which we employed only for validation purposes) or models that directly provide uncertainty predictions, interpreted as a measure of bias.
However, the choice of the classification can in principle enhance the flexibility of the protocol.
For instance, in scenarios where the training space includes sparsely sampled regions, a Bayesian Neural Network (BNN) may exhibit similarly high uncertainty both within those under-sampled areas and for points outside the training domain. 
If the goal is to filter out all samples with high uncertainty, a BNN will not differentiate between these two cases.
In contrast, a properly-selected classification model, such as logistic regression (which provides a linear boundary separating the two domains), can better define a region where the model should not be applied and another where predictions, regardless of their accuracy, are still feasible. 
This distinction can be particularly valuable in materials discovery, where testing materials within the same domain as previously studied samples can help avoid overlooking promising candidates.
%
Additionally, a BNN may fail to capture the transition between regions with sharply different underlying physics if it has only been trained on one of them.
As a result, such a model might underestimate uncertainties in regions that were not explored during training.
On the contrary, the methodology we propose would drop all the instances clearly coming from non-explored regions, as classifiers are trained not only on the regression training space but rather on the general purpose space.
%
Also, our approach can be seamlessly applied to any architecture, including modern graph equivariant neural networks, without the requirement to generate posterior distributions, as it would be necessary in BNN-based models.

%
%
%
%
A relevant example in materials discovery stems from a recently published study by Cerqueira \emph{et al.}~\cite{cerqueira2024sampling}, where the authors performed electron-phonon calculations of conventional superconductivity for $\sim7000$ materials; thus, over such database, they constructed a ML regression model applied to $\sim200,000$ out-of-the-box metallic materials for the prediction of the $T_{\mathrm{c}}$.
Among such $\sim200,000$ metallic materials examined, the authors focused on those with a ML-predicted critical temperature $T_{\mathrm{c}}>5\,\si{\kelvin}$; subsequently, a manual filtration based on single feature bounds was employed, resulting in the selection of 1032 materials for further analysis. Notably, 70\% of this subset exhibited a critical temperature surpassing $5\,\si{\kelvin}$; in contrast, within the initial training database, only 10\% of the materials had a critical temperature exceeding $5\,\si{\kelvin}$. 
With respect to that work, our approach, consisting of a systematic, automatized and multidimensional filtration, may achieve an increase of the aforementioned 70\%, resulting in a save of computational resources.

Another pertinent instance arises from the recent publication by Merchant \emph{et al.}~\cite{merchant2023scaling}, where the authors generated a database -- GNoME -- comprising nearly 400,000 new and stable materials. When having an already trained model on a specialized database for the prediction of a property of interest, our approach may reveal useful to filter out GNoME (or any other AI-generated database) materials for which such prediction would be unreliable.


%
For instance, the quality of data might vary based on the diverse synthesis methods employed, and data bias could be influenced by the experimenters' subjective choices in pursuing specific objectives.

%
%
%
In perspective, we imagine that such a tool can be used in combination with modern generative models (e.g., diffusion models), where the latter can hypothesize new possible and untested materials while the former judge reliability of predictions.

\section*{Data availability}
Datasets of this study are available in Zenodo (DOI:10.5281/zenodo.13686339) \cite{trezza_2024_13686339}.

\section*{Code availability}
The codes used to obtain the results of this study are publicly available at https://github.com/giotre/cobra.

\clearpage
\bibliographystyle{naturemag}

\end{document}